\documentclass[10pt,conference]{IEEEtran}

\usepackage{cite}
\usepackage{graphicx}
\usepackage{url}
\usepackage{xcolor}
\usepackage[hidelinks]{hyperref}
\usepackage{tikz}
\usetikzlibrary{arrows.meta,positioning}

\newcommand{\dhaani}{\textsc{Dhaani}}
\newcommand{\adfd}{ADFD}
\newcommand{\toolurl}{\url{https://shraddhasurana.github.io/dhaani/}}
\newcommand{\releaseurl}{\href{https://github.com/ShraddhaSurana/dhaani/releases/tag/v1.0.2}{Dhaani v1.0.2}}
\newcommand{\iparcartifact}{\href{https://github.com/ShraddhaSurana/iProg/tree/main/ijclr_iparc}{IPARC package}}
\newcommand{\migrationartifact}{\href{https://doi.org/10.5281/zenodo.22054047}{migration artifact archive}}
\newcommand{\demovideo}{\url{https://youtu.be/SShWwOMgJ9c}}

\title{Dhaani: A Data-Flow-Driven Tool for Greenfield Construction and Brownfield Modernisation}

\author{\IEEEauthorblockN{Shraddha Surana and Ashwin Srinivasan}
\IEEEauthorblockA{Department of Computer Science and Information Systems\\
BITS Pilani, K. K. Birla Goa Campus, Goa 403726, India\\
\{p20220031, ashwin\}@goa.bits-pilani.ac.in\\
ORCID: 0000-0002-3009-3178; 0000-0002-4911-0038\\
Corresponding author: Shraddha Surana}}

\begin{document}
\bstctlcite{IEEEexample:BSTcontrol}
\maketitle

\begin{abstract}
Large language models can generate plausible code quickly, but an end-to-end
prompt leaves developers with little control over decomposition, interfaces,
or the assumptions that must survive later revisions. We present \dhaani, a
free desktop application for greenfield program construction and brownfield
repository modernisation. It makes an annotated data-flow diagram (\adfd) the
inspectable contract between intent and implementation. In the greenfield
workflow, \emph{Build}, a user turns natural-language requirements into a typed graph,
reviews process specifications and pre/post-conditions, ratifies generated
modules, and proceeds in dependency order for Python or .NET (C\#). In the brownfield workflow,
\emph{Migrate} uses source inference to recover an \adfd{} from a Fortran
repository, performs target generation in bounded dependency groups, and
exposes source--target alignment,
generated code, errors, time, token use, and cost. The tool keeps code and
artifacts on the user's machine and supports cloud or locally hosted models.
Two scientific synthesis case studies produced executable systems where
matched low/no-code baselines failed or were substantially less accurate. On
50 open-source Fortran repositories, the migration workflow made all 382
fixed source-oracle probes runnable and passed 327 (85.6\%); two direct
translation baselines each passed 83 (21.7\% of planned probes). Pre-built
installers and usage instructions are freely available. Tool and usage guide:
\toolurl. Video: \demovideo.
\end{abstract}


\section{Motivation and Contribution}
Code-generating models can improve developer productivity~\cite{peng2023copilot},
yet developers often alternate between accelerated exploration and careful
verification~\cite{barke2023copilot}. The risk is amplified for scientific
software: domain assumptions span data acquisition, transformation, modelling,
validation, and reporting, while a plausible-looking port may silently change
array, state, or numerical semantics. LLM-generated translations have indeed
been shown to introduce non-trivial bugs~\cite{pan2024lost}.

\dhaani{} addresses the engineering problem between a natural-language goal
and trustworthy modules: making decomposition, data dependencies, and module
contracts visible before code is accepted. Its central artifact is an
\emph{annotated data-flow diagram} (\adfd): a directed graph of processes,
data stores, external entities, typed flows, per-process specifications, and
pre/post-conditions. Unlike a one-shot prompt or an ephemeral chat plan, the
same artifact is editable, saved with the project, and used to order work.
The approach extends structured-analysis DFDs~\cite{demarco1979structured}
from documentation to an active generation and migration boundary.

\begin{figure}[t]
  \centering
  \begin{tikzpicture}[
    font=\scriptsize,
    node distance=0.38cm and 0.28cm,
    wfbox/.style={draw=black!55, rounded corners=1.4pt, align=center,
      minimum height=0.70cm, text width=2.25cm, fill=black!2,
      inner xsep=2.5pt},
    contract/.style={wfbox, draw=teal!70!black, fill=teal!9,
      line width=0.75pt},
    action/.style={wfbox, draw=orange!75!black, fill=orange!10},
    flow/.style={-{Latex[length=1.6mm]}, line width=0.5pt},
    loop/.style={flow, dashed, draw=teal!70!black}
  ]
    \node[wfbox] (start) {Select a local Fortran\\repository and run options};
    \node[contract, right=of start] (source) {Build a static profile and\\infer the source ADFD};
    \node[wfbox, right=of source] (inspect) {Inspect or edit the ADFD\\at the optional checkpoint};
    \node[action, below=of inspect] (generate) {Generate Python in\\dependency-ordered groups};
    \node[action, left=of generate] (compare) {Infer the target ADFD,\\compare, and refine};
    \node[wfbox, left=of compare] (results) {Inspect generated files,\\diagnostics, and provenance};

    \draw[flow] (start) -- (source);
    \draw[flow] (source) -- (inspect);
    \draw[flow] (inspect) -- (generate);
    \draw[flow] (generate) -- (compare);
    \draw[flow] (compare) -- (results);
    \draw[loop] (compare.south) -- ++(0,-0.22) -|
      node[pos=0.28, below, font=\tiny] {remaining structural gaps} (generate.south);
  \end{tikzpicture}
  \caption{Brownfield migration workflow. The user selects a repository and
  run settings; \emph{Migrate} recovers an \adfd, generates the target in
  dependency groups, refines structural gaps, and exposes the saved result.}
  \label{fig:brownworkflow}
\end{figure}
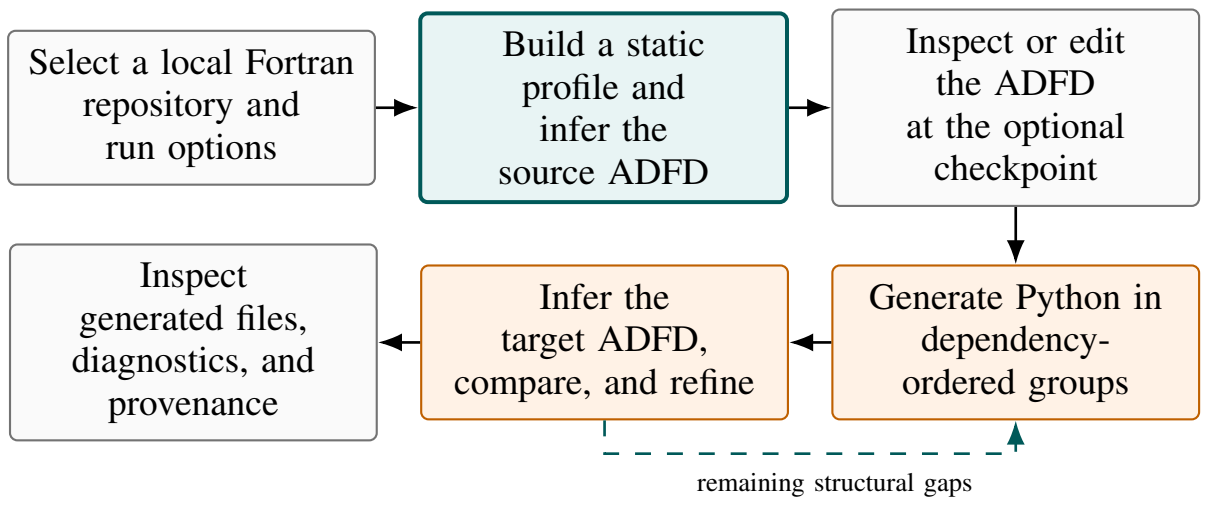

The intended users are (i) scientists and data practitioners constructing
greenfield procedural pipelines, (ii) software engineers who want a reviewable
LLM-assisted modularisation workflow, and (iii) maintainers modernising
brownfield repositories. The demonstration contributes:
\begin{itemize}
  \item an \adfd-centered development environment in which system structure,
        typed interfaces, conditions, and dependencies remain inspectable and
        editable as implementation proceeds;
  \item human ratification organized as bounded work
        against that explicit artifact rather than a monolithic response; and
  \item a local-first product surface that preserves provenance and separates
        operational diagnostics from independent behavioural evidence.
\end{itemize}
Prior studies used the tool for scientific-system construction, controlled
IPARC synthesis tasks, and repository migration~\cite{iprog,iparc,migrationpaper}.
Those studies provide technical evidence for particular uses of the \adfd.
This paper instead focuses on \dhaani{} as the usable realization of the
broader concept: an explicit \adfd{} is the primary working artifact through
which developers inspect, revise, generate, execute, and preserve decisions
about software. Greenfield construction and \emph{Migrate} are two supported
applications of this tool model. Figures~\ref{fig:greenworkflow}
and~\ref{fig:brownworkflow} give separate high-level usage paths for them.

\section{Tool Workflow}
\textbf{ADFD-centered development model.} \dhaani{} treats the \adfd{} also as the
primary development object. Each saved graph records processes, typed inputs and outputs, stores,
external entities, labeled flows, and editable pre/post-conditions. Declared
flows expose eligible work and carry ratified upstream interfaces into bounded
generation requests. For larger recovered graphs, strongly connected processes
remain together while the component graph is topologically ordered into bounded
groups. The following workflows show how these general tool capabilities are
applied when development starts from requirements or from an existing system.

\subsection{Greenfield (Build): From Requirements to Modular Code}
\begin{figure}[t]
  \centering
  \begin{tikzpicture}[
    font=\scriptsize,
    node distance=0.38cm and 0.28cm,
    wfbox/.style={draw=black!55, rounded corners=1.4pt, align=center,
      minimum height=0.70cm, text width=2.25cm, fill=black!2,
      inner xsep=2.5pt},
    contract/.style={wfbox, draw=teal!70!black, fill=teal!9,
      line width=0.75pt},
    action/.style={wfbox, draw=orange!75!black, fill=orange!10},
    flow/.style={-{Latex[length=1.6mm]}, line width=0.5pt},
    loop/.style={flow, dashed, draw=teal!70!black}
  ]
    \node[wfbox] (start) {Create a project by giving its description};
    \node[contract, right=of start] (adfd) {Generate and edit the\\ADFD and process contracts};
    \node[wfbox, right=of adfd] (select) {Select an eligible process\\from the workflow};
    \node[action, below=of select] (generate) {Generate code and\\execute the proposal};
    \node[action, left=of generate] (review) {Review explanation and\\ratify, revise, or reject};
    \node[wfbox, left=of review] (progress) {Save the module and\\update project progress};

    \draw[flow] (start) -- (adfd);
    \draw[flow] (adfd) -- (select);
    \draw[flow] (select) -- (generate);
    \draw[flow] (generate) -- (review);
    \draw[flow] (review) -- node[above, font=\tiny] {ratify} (progress);
    \draw[loop] (review.south) -- ++(0,-0.22) -|
      node[pos=0.28, below, font=\tiny] {revision feedback} (generate.south);
  \end{tikzpicture}
  \caption{Greenfield project workflow. The user establishes and edits an
  \adfd, then generates, executes, and ratifies eligible processes in
  dependency order until the modular program is complete.}
  \label{fig:greenworkflow}
\end{figure}
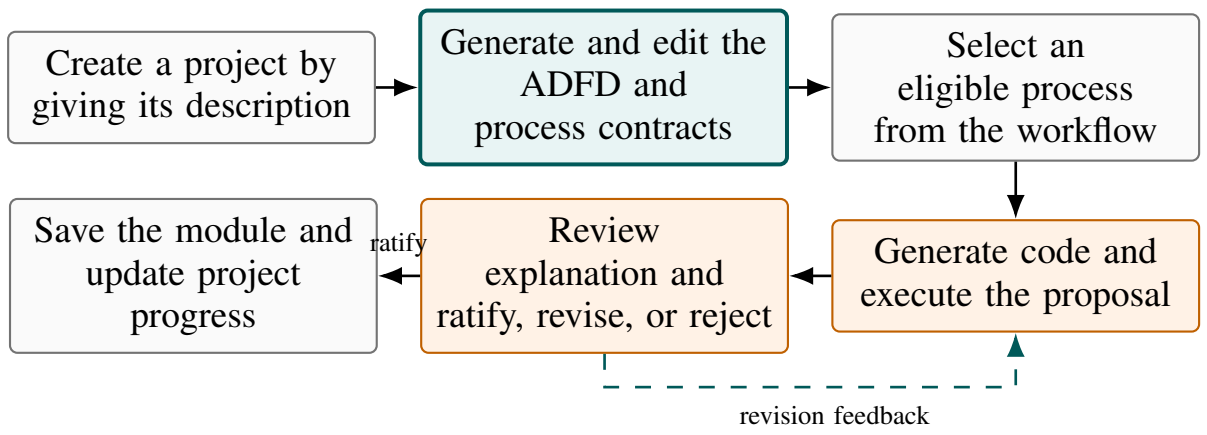

\begin{figure*}[t]
  \centering
  \begin{minipage}[t]{0.485\textwidth}
    \centering
    \includegraphics[width=\linewidth]{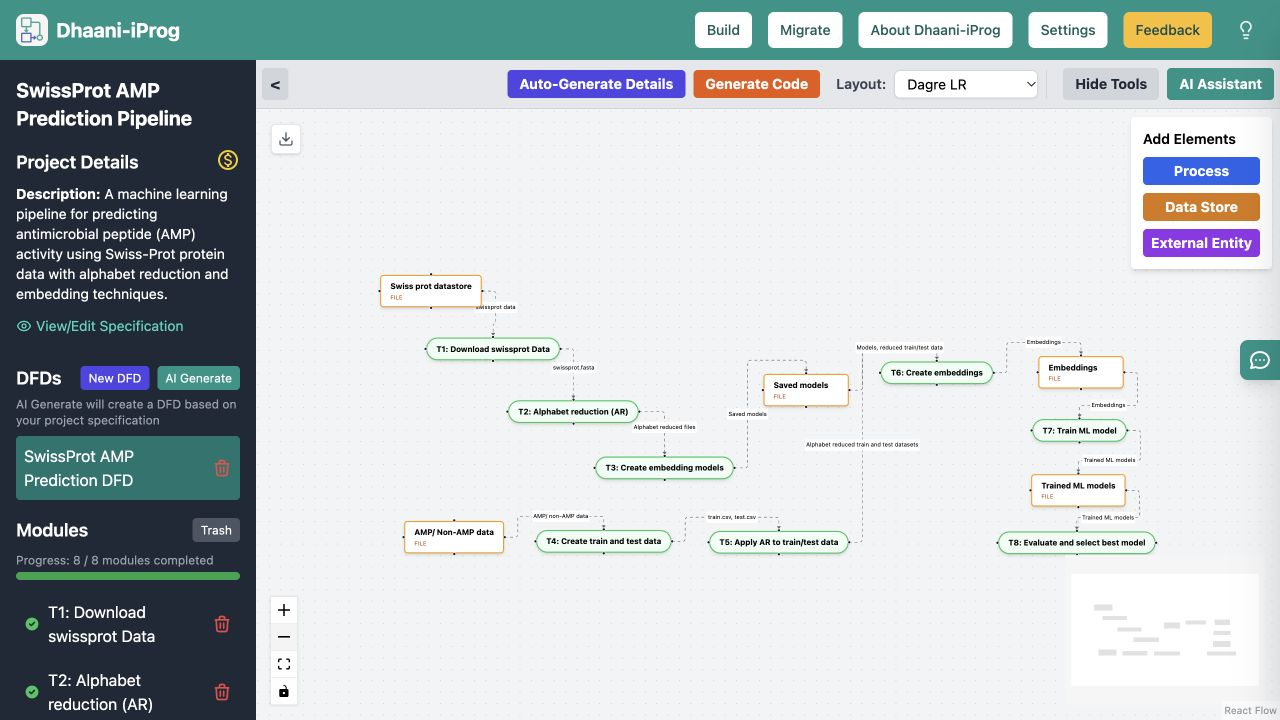}
    \centerline{\footnotesize (a) Editable SwissProt AMP \adfd.}
  \end{minipage}\hfill
  \begin{minipage}[t]{0.485\textwidth}
    \centering
    \includegraphics[width=\linewidth]{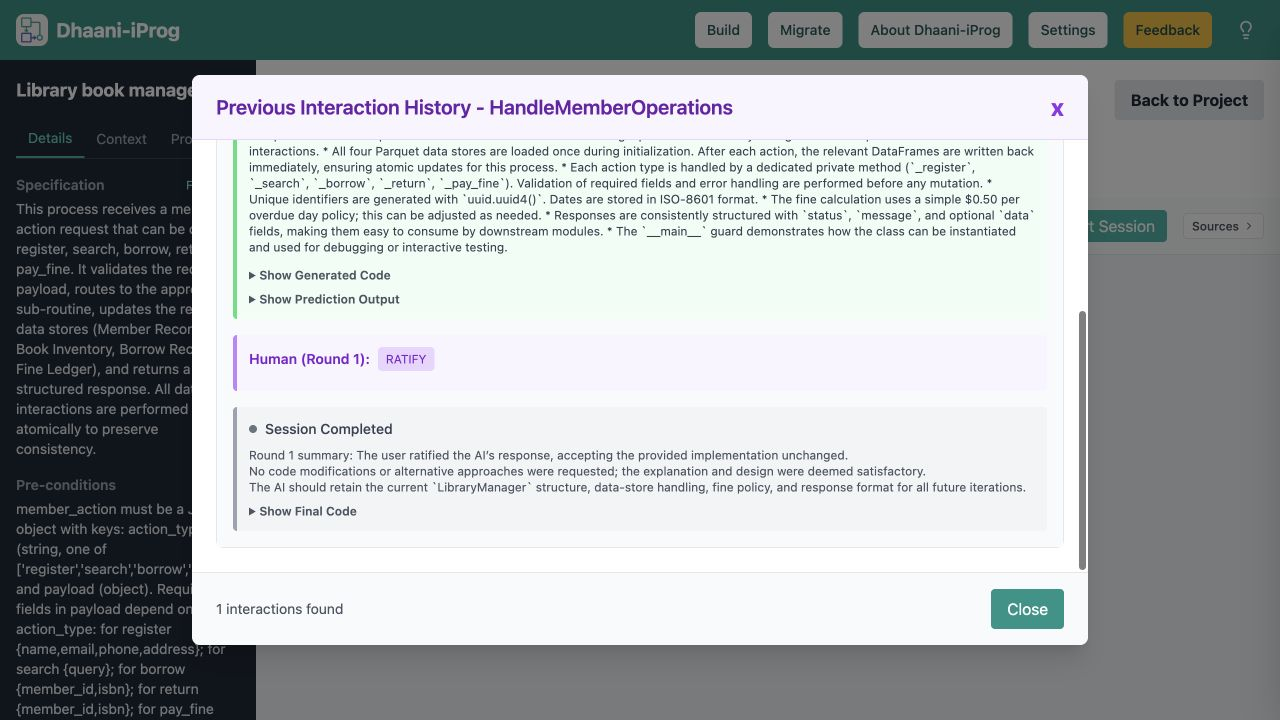}
    \centerline{\footnotesize (b) Library-management project module ratification record.}
  \end{minipage}
  \caption{Greenfield interfaces. The graph exposes the program decomposition
  and dependencies; a completed library-management process history preserves
  the model proposal, collapsible code/output views, the human \emph{Ratify}
  decision, and final code.}
  \label{fig:greenfield}
\end{figure*}

\textbf{From intent to an inspectable contract.} A user starts with a project
name and free-form specification, optionally supplies a dataset path, and
selects Python or .NET (C\#). Alternatively, one of three bundled scientific
projects can be cloned for immediate exploration. From the specification,
\dhaani{} can draft an \adfd{} of processes, data stores, external entities,
and labeled flows; the user can add, remove, reconnect, or rename these
elements before any code is generated. \emph{Auto-Generate Details} proposes
per-process specifications, typed inputs and outputs, and pre/post-conditions,
all of which remain editable. Optionally, user edits can receive an AI specification review.
Multiple diagrams can be maintained within a project, laid out automatically,
and exported as SVG. This makes decomposition and interface contracts visible rather than planning decisions hidden
inside an end-to-end prompt. Fig.~\ref{fig:greenfield}(a), for example, exposes the
complete path and branches from SwissProt retrieval to AMP-model evaluation
(one of the scientific projects).

\textbf{Generate, execute, and ratify one process at a time.} Selecting a
process in Build opens a bounded work item containing its responsibility, conditions,
incoming and outgoing data, and ratified upstream interfaces. The assistant
proposes code and an explanation, which the user assesses independently.
Users can edit and execute proposal code from the application, discuss it with the assistant, accept it
or provide focused revision feedback. Each round is retained as
ratify, revise, refute, or reject, preserving why the module was accepted instead of reducing
the decision to an ephemeral chat transcript. A successful ratification saves
the module implementation and advances project progress; a failed execution or
disputed explanation is localized to that module and can be corrected without
regenerating the whole system. The retained history in
Fig.~\ref{fig:greenfield}(b) makes the model proposal, optional code and output
views, human decision, and resulting completion state reviewable together.

\textbf{Assemble a dependency-consistent program.} The workflow view presents
all modules in data-dependency order and, for each, shows its specification,
conditions, data sources, dependencies, and interaction summary. Independent
modules can be completed first, while downstream work receives reviewed
interfaces and code from its predecessors. This combines the speed of model
generation with explicit checkpoints at the places where scientific
assumptions, array shapes, units, or interface choices can otherwise propagate
silently through a pipeline. The project sidebar and workflow-wide progress
indicator make incomplete or unratified modules visible at a glance.

\subsection{Brownfield: From a Repository to a New Implementation}
\textbf{Recover repository intent before generating a target.} The brownfield
workflow, \emph{Migrate}, currently migrates a local Fortran repository to
Python. The user may name the run, select single-pass inference for fewer model
calls or two-pass inference for a more detailed representation, and bound the
inference and refinement iterations. \dhaani{} first constructs a compact
static profile and performs \emph{Source ADFD Inference}. The recovered graph
makes processes, shared stores, external interactions, and cross-file data
flows reviewable as one repository-level contract. Coverage checks identify
important source elements not yet represented, and an optional checkpoint lets
the user inspect or edit the graph before generation. This addresses a central
migration risk: a plausible translation of each file can still omit repository
behaviour or break interfaces that only become apparent across files.

\textbf{Generate and refine with repository structure.} \emph{Target-Code
Generation} organizes work into bounded dependency groups and produces Python
modules in an order that preserves their relationships. It asks the model to
implement named responsibilities and interfaces expressed by the source
contract rather than to translate isolated files. \dhaani{} then infers an
\adfd{} from the generated target and compares it with the source graph.
Missing or additional processes, stores, and flows focus a bounded correction
pass on structural gaps instead of restarting the migration. This round-trip
check is an observable feedback
mechanism for repository coverage and consistency.

\begin{figure}[t]
  \centering
  \includegraphics[width=\columnwidth]{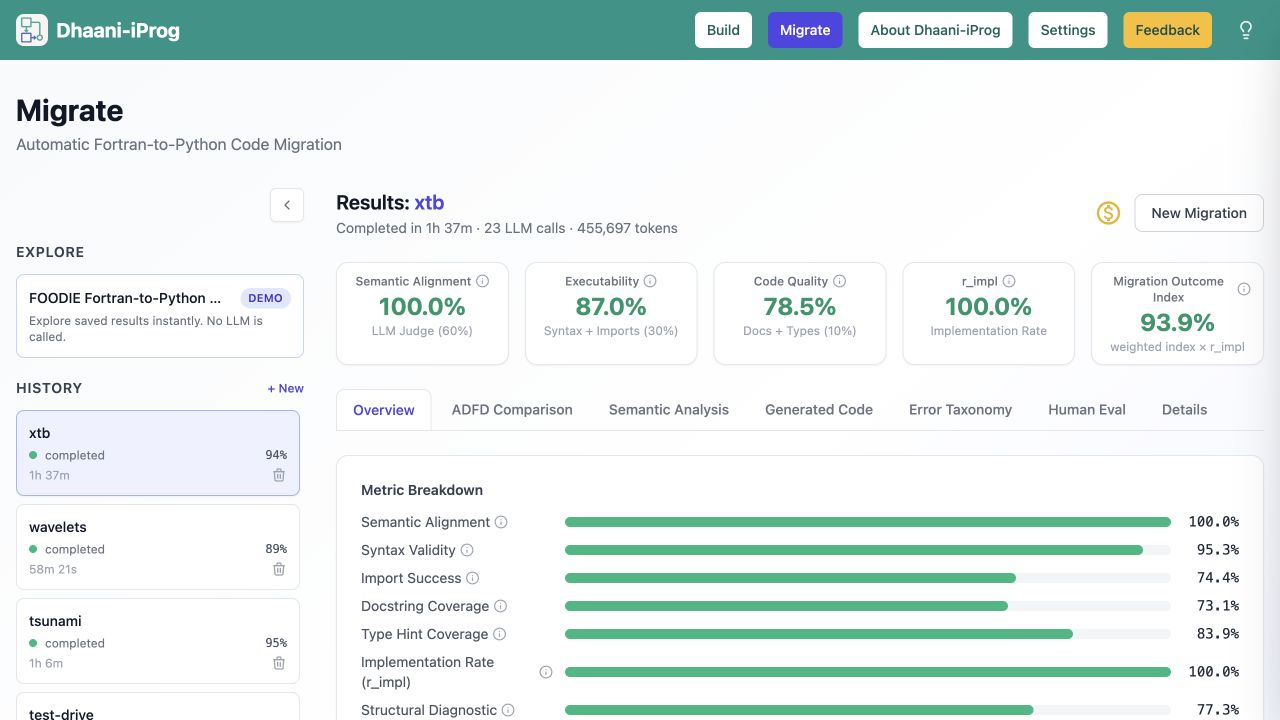}
  \caption{A completed Migrate run exposes repository history, generation
  telemetry, implementation checks, and drill-down views. Dashboard scores are
  operational diagnostics; the behavioural results in Section~III use
  independent source-oracle probes.}
  \label{fig:migrate}
\end{figure}

\textbf{Inspect the result and its provenance.} Live progress identifies the
active stage, while a persistent history makes completed runs immediately
reopenable. The results workspace (Fig.~\ref{fig:migrate}) provides overview,
source--target graph comparison, semantic alignment, generated-code, error
taxonomy, and execution-detail views. A maintainer can browse every generated
file, inspect unmatched graph elements and categorized diagnostics, and trace
per-stage time, model calls, tokens, and provider cost. These views separate
generation from assessment and help locate whether an incomplete result arose
from source inference, target generation, or refinement. The dashboard's
scores are operational diagnostics.

\subsection{Deployment, Privacy, and Model Choice}
The demonstration uses \releaseurl\footnote{Release:
\url{https://github.com/ShraddhaSurana/dhaani/releases/tag/v1.0.2}}. Its macOS,
Windows, and Linux installers
were uploaded on 10 October 2026. Projects, generated
code, diagrams, and interaction logs remain on the local machine. A bundled
interface and companion local service persist each project as a standalone
directory; a provider adapter isolates model-specific APIs.
Users may select a cloud provider or a local Ollama server; provider
credentials are entered through the local Settings panel and saved in a
local application configuration file, separate from individual project
directories. When a cloud model is selected,
relevant specification or code context is transmitted to that provider; a
local model avoids this transfer. The application is free to download and use,
while external model charges remain the user's responsibility. The bundled
Star Formation, Iris, and SwissProt projects give users an immediate exploration point
through the editor. Users can install the public distribution from
\toolurl{} and inspect seeded projects.  Completed Migrate runs are available from the \migrationartifact{}
\footnote{Migration artifact:
\url{https://doi.org/10.5281/zenodo.22054047}}.

The tool records tokens, model-specific cost, and elapsed time, exposes
fewer-call and higher-quality inference modes, and can run compatible local
models.

\section{Validation}
We report complementary greenfield and brownfield studies. They validate the
research workflows implemented by the tool; they are not a usability
evaluation of every UI control.

\textbf{Greenfield evidence.} Starting only from natural-language problem
descriptions, one engineer used the forward workflow to reconstruct two
published data-analysis systems: astrophysical regression over 76,455
galaxies and antimicrobial-peptide classification. The same model and a
matched interaction budget were used for free-form low-code baselines, while
no-code baselines received one end-to-end prompt. \dhaani{} produced a 97.6\%
accurate peptide classifier while the low/no-code systems were non-functional;
for the astrophysics targets its errors were 5--12$\times$ lower than the
functional low-code baselines. A separate controlled study used the same
structured approach to IPARC's 600 tasks, spanning sequence, selection, and
iteration categories. This was the first reported success on the benchmark
across all categories~\cite{iparc}. Full scientific protocols and manual comparisons
appear in the method study~\cite{iprog}.

\textbf{Brownfield evidence.} We selected 50 open-source Fortran
repositories larger than 1,000 lines, spanning 20 low-, 20 medium-, and 10
complex-tier projects and without an existing public Python translation.
Before examining generated ports, we fixed 382 probes from source behaviour.
Each probe executed the pinned Fortran routine (directly or through
\texttt{f2py}~\cite{peterson2009f2py}) and the generated Python on identical
inputs with probe-specific exact or numerical comparison. \dhaani{} Migrate made
every planned target runnable and passed 327; 40 of 50 repositories passed all
their probes. Each direct baseline passed 83 overall, largely because only 99
and 98 planned behaviours were runnable. When a baseline target did run, its
conditional agreement (83.8\% and 84.7\%) was close to \dhaani{} Migrate's 85.6\%;
the improvement is therefore primarily repository completeness, not inflated
per-routine accuracy. Two no-\adfd{} ablations passed 11.0\% and 3.7\% of the
same denominator, supporting the representation-and-workflow
mechanism~\cite{migrationpaper}.

The bundled scientific projects provide runnable representative workflows.
Supporting materials for the prior IPARC and migration evaluations are
available in the \iparcartifact{}\footnote{IPARC package:
\url{https://github.com/ShraddhaSurana/iProg/tree/main/ijclr_iparc}} and
\migrationartifact{}, respectively.

\textbf{Current scope and expansion.} The end-to-end greenfield evidence covers
two scientific systems and is complemented by the 600-task IPARC challenge,
which broadens coverage across control structures. Migrate currently
specializes in Fortran-to-Python; the evaluation uses targeted source-oracle
probes rather than exhaustive whole-program verification. The remaining
migration failures point to clear opportunities for stronger language-aware
analysis in state-heavy, macro-rich, and dependency-intensive repositories.
The product has since expanded greenfield generation to .NET 8/C\#, while
database integration is available as an early-stage capability; systematic
evaluation of these newer capabilities is future work. As with other
model-assisted tools, outputs vary with the selected model and configuration.
The current evidence is therefore strongest for procedural and scientific
software, with product and evaluation coverage expanding to additional
languages and application classes.

\section{Related Tools and Demonstration}
Bindings such as \texttt{f2py} retain the original Fortran binary, while
FABLE and LFortran use rule- or compiler-oriented paths~\cite{grosse2012fable,lfortran2023}.
Repository-level LLM systems instead use program analysis or intermediate
translations: AlphaTrans uses static analysis to decompose Java repositories
and translates fragments in reverse call order~\cite{ibrahimzada2025alphatrans},
while InterTrans routes code through
intermediate programming languages~\cite{macedo2025intertrans}. For greenfield
work, GUIDE decomposes GUI descriptions into fine-grained requirements and
editable Figma prototypes~\cite{kolthoff2025guide}; CodePlan plans repository-wide
edits using dependency and change-impact analyses~\cite{bairi2024codeplan}.
\dhaani's distinctive pivot is a user-editable, typed data-flow and contract
representation that governs bounded software-development work. The current
tool applies this model to modular greenfield generation and brownfield
modernisation.

The accompanying demonstration follows one scientific example: (1) clone the
bundled Star Formation, Iris, or SwissProt project; (2) inspect and edit its \adfd; (3) open a process, review
pre/post-conditions, run a proposal, and ratify it; (4) inspect the
dependency-ordered workflow; and (5) open a completed Migrate run to compare
source/target diagrams, generated files, diagnostics, and resource use. This
path exposes the central interaction within minutes while avoiding a long
live migration during the session.

\section{Conclusion}
\dhaani{} makes the \adfd{} a persistent, inspectable development artifact
between software intent or an existing implementation and generated modules.
It exposes decomposition, contracts, dependencies, ratification, and
provenance as bounded work rather than leaving them inside model responses.
Dhaani applies this ADFD-centred development approach to greenfield
construction and brownfield modernisation;
its strongest evidence is for scientific procedural software across two
end-to-end systems, controlled IPARC tasks, and 50 Fortran-to-Python
repositories. Future work will evaluate the current .NET support and
early-stage database integration, add language-specific analyzers and
generators, and broaden frontend application coverage.

\bibliographystyle{IEEEtran}
\bibliography{references}

\end{document}